\documentclass[sigconf]{acmart}

\usepackage{tabularx}

\usepackage{graphicx}
\usepackage{subfigure}
\usepackage{booktabs} 
\usepackage{enumitem}
\usepackage{multirow}
\usepackage{pdfrender}
\usepackage{array}
\usepackage{hyperref}
\usepackage{makecell}
\usepackage{amsmath}
\usepackage{gensymb}
\usepackage{amsfonts}
\usepackage{mathrsfs}
\usepackage{balance}
\usepackage{xcolor}
\theoremstyle{definition}

\newcommand{\RNum}[1]{\uppercase\expandafter{\romannumeral #1\relax}}
\usepackage[linesnumbered,ruled,vlined]{algorithm2e}

\SetCommentSty{mycommfont}

\SetKwInput{KwInput}{Input}                
\SetKwInput{KwOutput}{Output} 
\SetKwFunction{KwFunction}{Function}

\AtBeginDocument{%
  \providecommand\BibTeX{{%
    \normalfont B\kern-0.5em{\scshape i\kern-0.25em b}\kern-0.8em\TeX}}}
\copyrightyear{2023}
\acmYear{2023}
\setcopyright{acmlicensed}\acmConference[WiSec '23]{Proceedings of the 16th ACM Conference on Security and Privacy in Wireless and Mobile Networks}{May 29-June 1, 2023}{Guildford, United Kingdom}
\acmBooktitle{Proceedings of the 16th ACM Conference on Security and Privacy in Wireless and Mobile Networks (WiSec '23), May 29-June 1, 2023, Guildford, United Kingdom}
\acmPrice{15.00}
\acmDOI{10.1145/3558482.3590195}
\acmISBN{978-1-4503-9859-6/23/05}

\usepackage[T1]{fontenc}
\usepackage[utf8]{inputenc}

\begin{document}

\title{
    HoneyIoT: Adaptive High-Interaction Honeypot for IoT Devices Through Reinforcement Learning
}

\author{Chongqi Guan}
\email{cxg5270@psu.edu}
\affiliation{%
  \institution{The Pennsylvania State University}
  \city{University Park}
  \state{PA}
  \country{USA}
}

\author{Heting Liu}
\email{hxl476@psu.edu}
\affiliation{%
  \institution{The Pennsylvania State University}
  \city{University Park}
  \state{PA}
  \country{USA}
}

\author{Guohong Cao}
\email{gxc27@psu.edu}
\affiliation{%
  \institution{The Pennsylvania State University}
  \city{University Park}
  \state{PA}
  \country{USA}
}

\author{Sencun Zhu}
\email{sxz16@psu.edu}
\affiliation{%
  \institution{The Pennsylvania State University}
  \city{University Park}
  \state{PA}
  \country{USA}
}

\author{Thomas La Porta}
\email{tfl12@psu.edu}
\affiliation{%
  \institution{The Pennsylvania State University}
  \city{University Park}
  \state{PA}
  \country{USA}
}



\begin{abstract}
As IoT devices are becoming widely deployed, there exist many threats to IoT-based systems due to their inherent vulnerabilities.  
One effective approach to improving IoT security is to deploy IoT honeypot systems, which can collect attack information and reveal the methods and strategies used by attackers. However, building high-interaction IoT honeypots is challenging due to the heterogeneity of IoT devices. Vulnerabilities in IoT devices typically depend on specific device types or firmware versions, which encourages attackers to perform pre-attack checks to gather device information before launching attacks. Moreover, conventional honeypots are easily detected because their replying logic differs from that of the IoT devices they try to mimic.
To address these problems, we develop an adaptive high-interaction honeypot for IoT devices, called {\em HoneyIoT}.   We first build a real device based attack trace collection system to learn how attackers interact with IoT devices.  We then model the attack behavior through markov decision process and leverage reinforcement learning techniques to learn the best responses to engage attackers based on the attack trace. 
We also use differential analysis techniques to mutate response values in some fields to generate high-fidelity responses.
HoneyIoT has been deployed on the public Internet. Experimental results show that HoneyIoT can effectively bypass the pre-attack checks and mislead the attackers into uploading malware. 
Furthermore, HoneyIoT is covert against widely used reconnaissance and honeypot detection tools. 

\end{abstract}

\maketitle

\vspace{-0.3cm}

\section{Introduction}
\label{sec:introduction}

Internet of Things (IoT) are being widely deployed in a variety of consumer, enterprise, and military settings such as home automation, smart manufacturing, autonomous driving, smart city, and military operations.  However, there exist many threats to IoT devices due to their inherent vulnerabilities \cite{neshenko2019vulnerabilitysurvey,Sok_SP19,Circle_Usenix21,IoTVul_CCS22}, which are caused by outdated or broken security modules, improper patch management, insufficient access control, and inadequate physical security. 
Many recent attacks \cite{Mirai,Mozi} utilize these security vulnerabilities to compromise and infect IoT devices. Therefore, 
there is a pressing need to understand the dynamic threat landscape for IoT devices in order to improve their overall security.

Honeypot is a valuable security tool that has been widely used by security practitioners to gain insight into the dynamic threat landscape.  
They are decoy systems designed to attract, engage and deceive potential attackers through vulnerable services in a monitored and controlled environment. Typically, they consist of virtual systems that closely mimic real production environments to effectively engage attackers. A successful honeypot can efficiently attract attackers through different vulnerabilities, evade reconnaissance and honeypot detection tools, provide seemingly genuine responses, and collect any attack traces left by the attackers for future analysis. As such, honeypots have the potential to be applied to improve IoT security.

Applying honeypot technology to the IoT domain presents significant challenges. This is primarily due to the heterogeneity of IoT devices, which means that vulnerabilities can vary significantly depending on the specific device brand, model, or firmware version. 
As a result, attackers typically perform various pre-attack checks to gather information about the target device before launching attacks. 
For instance, Nmap \cite{Nmap} is a widely used reconnaissance tool that can perform port scans and identify devices based on probing results. Additionally, attackers can use honeypot detection tools such as HoneyScore \cite{shodanHoneyscore} or HoneypotHunter \cite{HoneyHunter} to determine if a device is a honeypot during their pre-attack checks. Traditional IoT honeypots \cite{pa2015iotpot} are often limited in terms of interactions due to their fixed replying logic, which makes it difficult for them to deceive attackers during pre-attack checks.

To address these issues, we first need to learn how an attacker interacts with real IoT devices. In order to model the attacker behavior, we collect attack traces by exposing vulnerable IoT devices to attackers directly. The collected attack traces contain the attack behavioral information such as what IoT device the attacker is targeting, when they send specific pre-attack checks, and what responses can better lead to vulnerability exploitation or malware uploads. Since the attack trace is too complicated to extract meaning information purely based on heuristic based method or human analysis, we model the attack behavior through markov decision process (MDP) and leverage reinforcement learning techniques to learn the best responses to engage the attackers based on the collected trace. 
In addition, differential analysis techniques are used to mutate the response values in certain fields (e.g., date, time, sensor values, etc.) in order to generate more authentic and convincing responses.

\begin{figure}[t!]
    \centering
    \includegraphics[width=8.5cm]{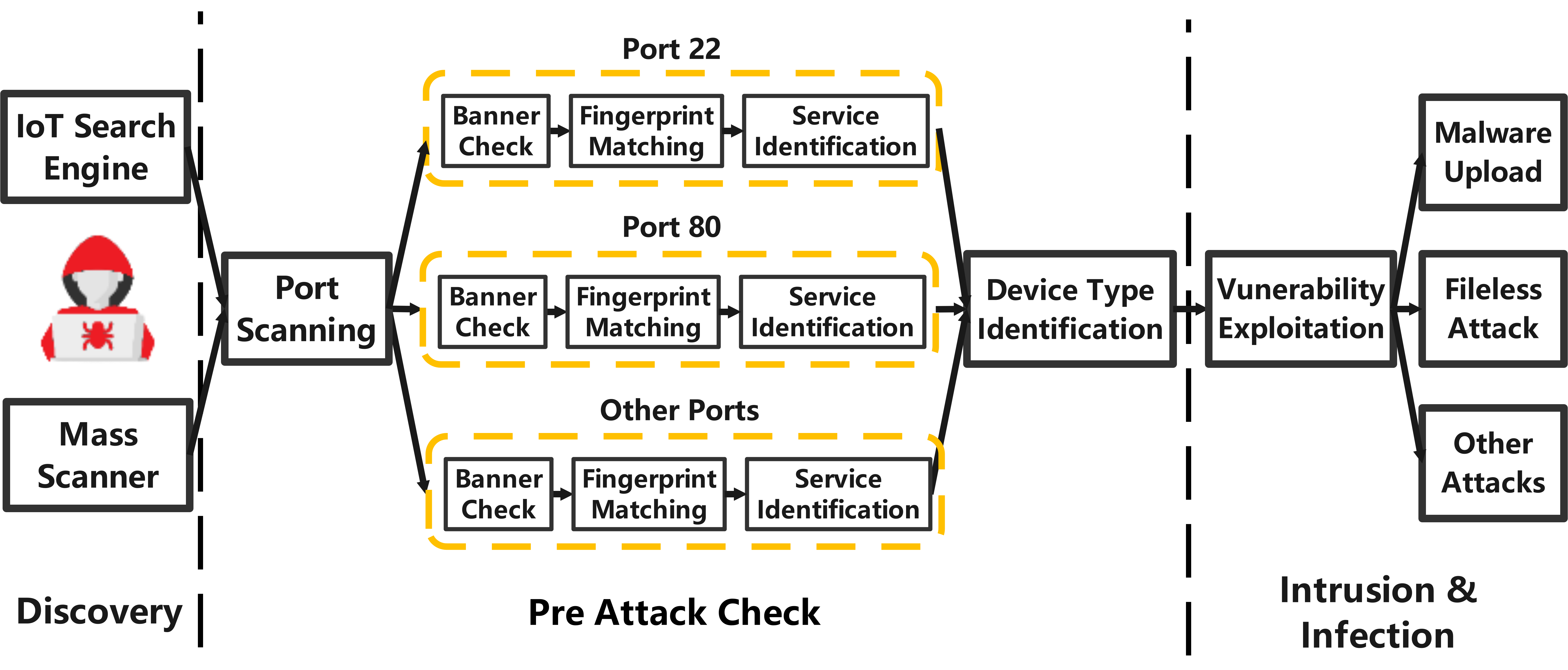}
    \vspace*{-0.25in}
    \caption{Attacks against IoT devices}
    \label{fig:AttackModel}
    \vspace*{-0.2in}
\end{figure}
The main contributions of this paper are as follows:
\begin{itemize}
    \item We propose HoneyIoT, an adaptive high-interaction honeypot for IoT devices through reinforcement learning.
    \item We develop techniques to model the interactions between attackers and honeypot as a markov decision process 
    base on the collected attack trace and leverage reinforcement learning to engage attackers.
    \item We identify the mutation fields in the response through differential analysis and update them in real time to provide high fidelity responses.
    \item We evaluate the effectiveness and robustness of HoneyIoT by deploying the system on the public Internet. Our evaluation results show that HoneyIoT is covert against widely used reconnaissance and honeypot detection tools, and it can  mislead attackers into uploading malware. 
\end{itemize}

\section{Background}
\label{sec:background}

In this section, we provide background information about the evolving threat landscape for IoT devices 
and explain why we need to build a more deceptive and interactive IoT honeypot.

IoT devices have long been valuable targets to attackers due to their inherent vulnerabilities \cite{neshenko2019vulnerabilitysurvey,Sok_SP19,Circle_Usenix21,IoTVul_CCS22}. For example, the storm of Mirai botnet \cite{Mirai} has overwhelmed several high-profile targets since late 2016. The original Mirai botnet compromises IoT devices through brute-force login against the telnet port.
After a successful login, Mirai bots try to perform a series of operations to infect target devices and corral them into a botnet. The Mirai botnet was able to infect 
hundreds of thousands of IoT devices, which were used to perform DDos attacks against different targets. However, with the upgrading of IoT devices over recent years, 
simple brute-force login attacks against telnet ports are no longer effective. Some IoT device manufactures choose to use random default passwords to mitigate password cracking, and others choose to disable telnet and ssh services to avoid being discovered by the Mirai botnet.

Meanwhile, the attack against IoT devices is evolving. Instead of password cracking, attackers nowadays are targeting various types of vulnerabilities on IoT devices. For example, by exploiting the Remote Code Execution (RCE) vulnerabilities, attackers can inject malicious code to corral the target devices into their botnet. Since the vulnerabilities usually depend on the type, brand, and model of the target IoT devices, attackers nowadays tend to perform several pre-attack checks on the device before injecting malicious code to increase the attack success rate.

Figure \ref{fig:AttackModel} shows a typical attack process against IoT devices. The attacker first uses some reconnaissance tools such as Mass scan \cite{masscan} or IoT search engine \cite{shodan} to locate the victim device. 
Then, the attacker scans and probes the open ports of the target device to gather more information. 
The responses from the remote host can be used to match known fingerprints of existing honeypots \cite{Survey_Vuner, HoneypotDetection1,HoneypotDetection2,HoneypotDetection3}.
For example, many open-source honeypots offer a limited set of hard-coded banners or static http responses which can be used as fingerprints to identify the remote hosts. 
During these pre-attack checks, if the attacker observes any honeypot fingerprint, or finds out inconsistency between the simulated device and the provided service, he will suspect that he is interacting with a honeypot. 
The attacker will either evade these honeypots by blacklisting their IP addresses, 
or takes down these honeypots through DDos attacks. 
The responses can also be used to identify the service running on the remote host, which can further be leveraged to pinpoint the type of the victim IoT device \cite{fingerprintIoT3USENIX}. 
For example, if the port scan results indicate that the target is providing a video streaming service through real time streaming protocol (RTSP) on port 554, and a Web server for camera control on port 80, 
the target device is most likely an IoT camera. After identifying the remote host, the attacker can speculate 
the types of vulnerabilities existing on the target IoT device 
and then launch the exploitation attack. If the attack is successful, the attacker 
may launch various types of follow-up attacks such as uploading malware or paralyzing the device.

Honeypots have been used to defend against attacks on IoT devices. Conventional IoT honeypots mainly focus on emulating specific protocols such as telnet or ssh \cite{pa2015iotpot,RLHoney1}. The attacker may notice that certain service is missing and suspect that he is not interacting with a real IoT device. 
In addition, the honeypot only provides limited level of interaction to attackers with some fixed replying logic (i.e., fixed answer to various requests). Such behavioral fingerprints may have been recorded by the scanners \cite{Nmap, masscan} used by the attackers to identify the honeypots. 
Therefore, conventional IoT honeypots are not effective against the latest attackers. There is a strong need to build adaptive high-interaction IoT honeypot which can interact with the attackers, bypass their pre-attack checks,
and mislead them to upload their malicious codes.  
\section{Attack Trace Collection}
\label{sec:AttackTrace}

In this section, we present a system to collect attack traces, which can help us learn how the attackers interact with real IoT devices. We also show some analysis results based on the attack traces and generate attack graphs for IoT devices.

\subsection{Real-Device Based Attack Trace Collection}

\begin{figure}[tbp]
    \vspace*{-0.1in}
    \centering
    \includegraphics[width=8.5cm]{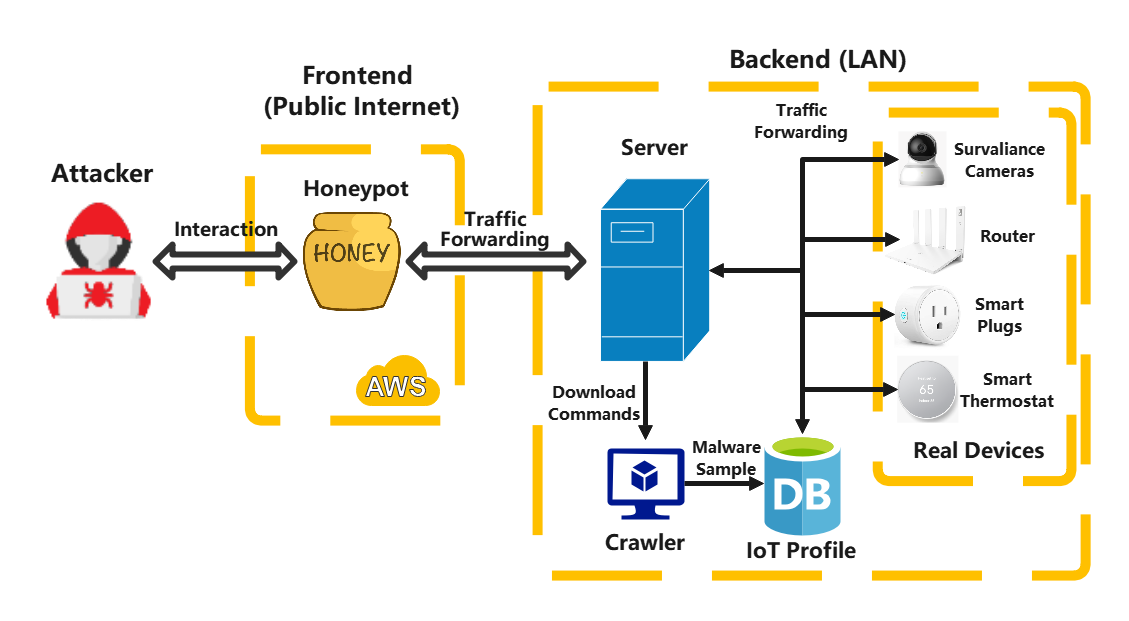}
    \vspace*{-0.35in}
    \caption{Attack trace collection based on real IoT devices}
    \label{fig:VPNsystem}
    \vspace*{-0.15in}
\end{figure}

We build a system to collect attack traces based on the interactions between attackers and real IoT devices.
As shown in Figure \ref{fig:VPNsystem}, the system consists of a frontend virtual machine running on AWS, a backend server for traffic forwarding and preliminary traffic analysis, and a few IoT devices including various models of IoT cameras, routers and smart plugs. 
The detailed list of IoT devices and their corresponding vulnerabilities are shown in Table~\ref{tab:device}.
The system interacts with the attacker by forwarding the received packet to one of the IoT devices 
to learn what the attacker will do next. Based on the attacker's request, the corresponding IoT device 
sends the necessary files or responses so that the attacker can continue to interact with the corresponding IoT device. This process continues until the attacker uploads some exploit code or stops interacting with the IoT devices.
Our system maintains the log traces and may have to be rebooted in some cases to recover from the attacks. Then, a new cycle starts which may select a different IoT device for a different attacker.  By doing this, our system can obtain different attacker traces, targeting different kinds of IoT devices, or targeting different kinds of protocols.
We use the open source project SysFlow \cite{Sysflow} to monitor the traffic between the attacker and the IoT devices and perform event-driven analysis to classify requests. We filter out any commands containing download instructions such as Wget or Curl, and forward them to a crawler in a sandbox to automatically collect malware from the attacker's control and command server.

\begin{figure*} [t]
   \centering 
   \subfigure[]{ 
     \label{fig:AttackGraphNC220}  
     \includegraphics[width=8.25cm,height = 4.75cm]{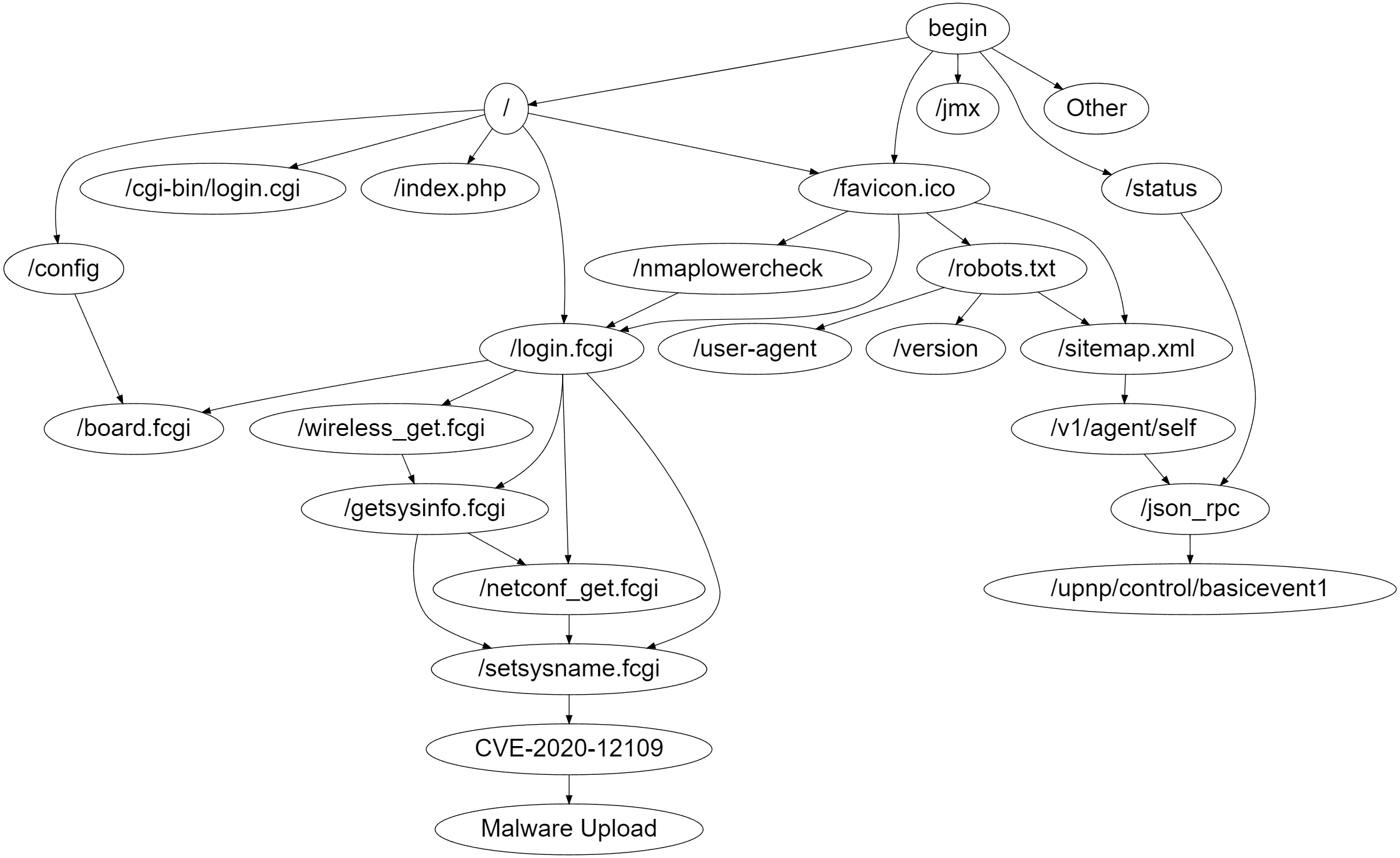}} 
   \subfigure[]{ 
     \label{fig:AttackGraphReolink} 
     \includegraphics[width=8.75cm,height = 4.75cm]{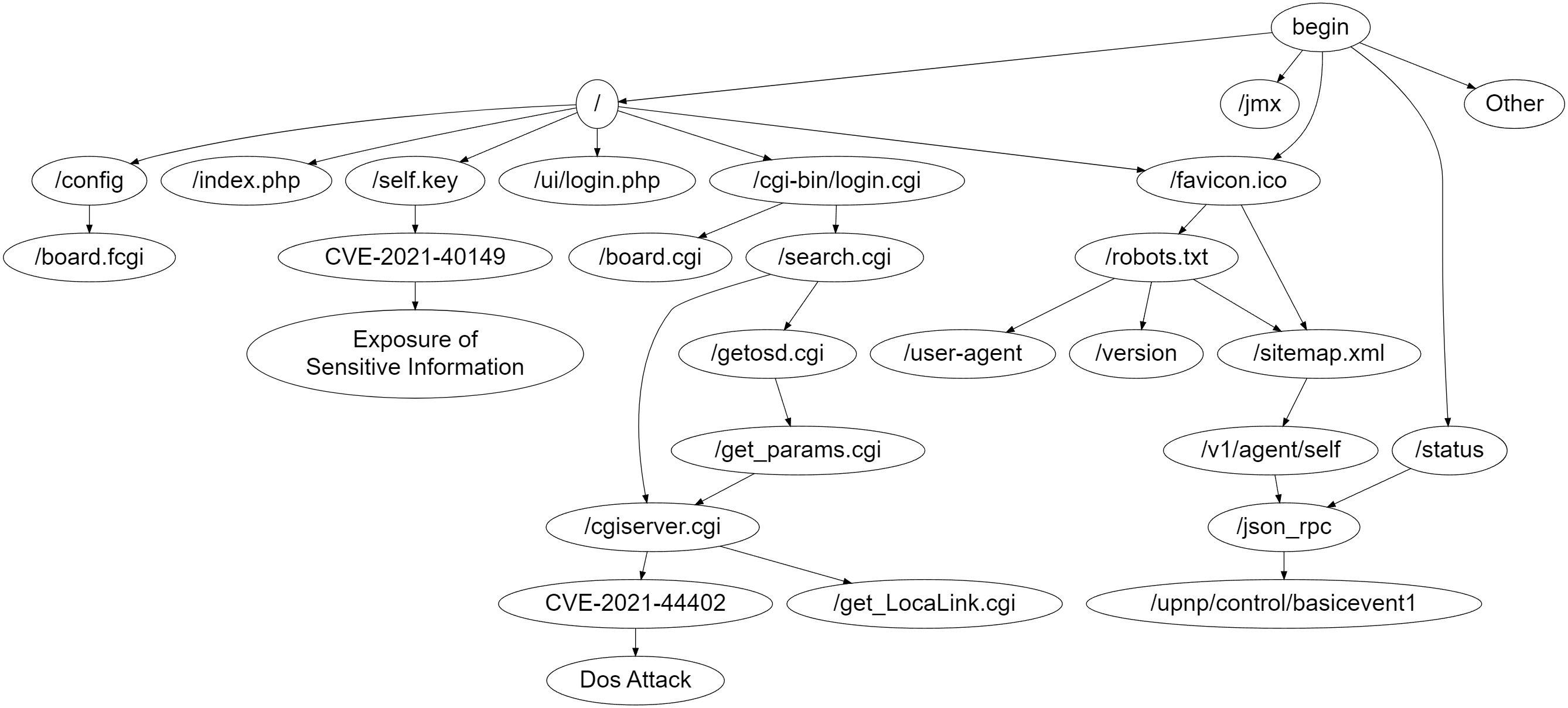}} 
   \vspace{-0.2in}
   \caption{Partial Attack graph against HTTP port: (a) TPlink TC220 camera (b) Reolink Camera}
   \label{fig:AttackGraph} 
   \vspace{-0.1in}
 \end{figure*}

\subsection{Preliminary Analysis on Attack Traces}
\label{sec:AttackTraceAnalysis}

Our attack trace collection system was deployed on AWS from June 17 to September 1, 2022. During this period, the attack trace was collected to model the attack behavior and gain insights on how to effectively engage with attackers.
An initial analysis of the trace indicates that it encompasses all known remotely exploitable CVEs associated with these IoT devices.
In order to better understand the attacker behavior against IoT devices through the attack traces, 
we generate attack graphs for IoT devices based on the interactions between the attackers and the IoT devices.
Figure \ref{fig:AttackGraph} shows two attack graphs against TPlink NC220 cameras and Reolink cameras over HTTP ports. As the whole graph is way too big, 
we only show a partial attack graph emphasizing specific vulnerabilities exploited and the attack behavior.

In the attack graph, a node represents the attacker's action such as probing a directory, accessing a resource, exploiting a certain vulnerability or uploading a malware. 
For example, the node with `/' indicates that the attacker probes the root directory. 
The node with `/favicon.ico' indicates that the attacker tries to access the favicon file 
which is usually a small icon indicating the device type or manufacturer.
The node with `CVE-2020-12109' means that the attacker exploits a specific vulnerability with Common Vulnerabilities and Exposures (CVE) ID 2020-12109, i.e., a publicly disclosed IoT security flaw on TPlink web camera 
where no format check is enforced when the attacker sends malicious HTTP request through `/setsysname.fcgi'. 
The edges connecting two nodes indicate that some attackers have taken another action after receiving the previous response 
from the IoT device. 

\begin{table}[]
\resizebox{.42\textwidth}{!}
{
\begin{tabular}{|l|l|l|l|}
\hline
Device Model & Manufacture & Device Type & Vulnerability ID    \\ \hline
NC220        & TP-Link     & Camera      & CVE-2020-12109, etc \\ \hline
RLC-410W     & Reolink     & Camera      & CVE-2021-44402, etc \\ \hline
E1 Zoom      & Reolink     & Camera      & CVE-2021-40149      \\ \hline
Home         & YI          & Camera      & CVE-2018-3928, etc  \\ \hline
DS-2CD2183G  & Hikvision   & Camera      & CVE-2021-36260      \\ \hline
Insight      & Wemo        & Smart Plug  & CVE-2018-6692       \\ \hline
Mini         & Wemo        & Smart Plug  & CVE-2018-6692       \\ \hline
HS103-P4     & TPlink      & Smart Plug  & CVE-2019-15745      \\ \hline
ISP5         & iHome       & Smart Plug  & RCE \footnote{\label{note1} This Vulnerability does not have a Common Vulnerabilities and Exposures Number yet}      \\ \hline
ISP6         & iHome       & Smart Plug  & RCE \footnotemark[1]      \\ \hline
VMB3000      & Netgear     & Router      & CVE-2019-3949, etc  \\ \hline
DGN2220      & Netgear     & Router      & CVE-2020-35577, etc \\ \hline
TL-WR840N    & TP-Link     & Router      & CVE-2018-14336, etc \\ \hline
DIR-3040     & D-Link      & Router      & CVE-2021-21819, etc     \\ \hline
WS5200       & Huawei      & Router      & CVE-2019-5268, etc  \\ \hline
WS7200       & Huawei      & Router      & N/A                 \\ \hline

\end{tabular}
}
\vspace*{0.1in} 
\caption{IoT devices used in our attack trace collection system}
\vspace*{-0.5in}  
\label{tab:device}
\end{table}

From the attack graph, we can see that the attacker conducts various types of pre-attack checks to gather information from the remote host before launching attacks. The attacker may choose different follow up attacks based on the responses from the IoT devices.
For example, as shown in Figure \ref{fig:AttackGraphNC220}, some of the attackers first access the favicon file to identify that this is a TPlink camera, by matching the MD5 hash of favicon or by analyzing its image. 
Then, they decide to exploit the `CVE-2020-12109' vulnerability by sending various requests. 
On the other hand, if the attacker notices that the remote host is not a TPlink camera (the MD5 has does not match), he may not proceed with follow up attacks. 
As shown in Figure \ref{fig:AttackGraphReolink}, the attack trace collected in the Reolink camera does not have such attack behavior because the favicon of Reolink camera is different from the favicon of TPlink camera.

The collected attack trace contains valuable attacker behavioral information which can be used to answer questions such as what IoT device the attacker is targeting, when the attacker sends specific pre-attack checks, and which response can better lead to a vulnerability exploitation or malware upload. The answers to these questions are critical to build high-interaction honeypots for IoT devices. 
In order to effectively learn the attack behavior through these collected attack traces, HoneyIoT leverages reinforcement learning techniques.

\section{Adaptive High-Interaction Honeypot for IoT Devices}
\label{sec:HoneyIoT}

In this section, we introduce HoneyIoT, an adaptive high-interaction honeypot for IoT devices. We first formulate the interactions between honeypots and attackers as a Markov Decision Process (MDP), and then leverage reinforcement learning to build an agent which can adaptively interact with the attacker by selecting proper responses. We also propose a differential analysis based content mutation method to effectively update some fields of the response at run time. At last, we present the general system design of HoneyIoT. 

\subsection{Problem Formulation} 
After taking a close look at the collected attack traces, we notice that there is a strong correlation between the attacker's requests and the responses from the IoT devices. As discussed in Section \ref{sec:AttackTraceAnalysis}, some of the attacker's request packets look for the type and model information of the target device, and others try to collect the service and protocol information of that device. These packets together can be classified as "pre-attack checks",  which are used to identify the target IoT device. Since attackers performing different pre-attack checks may be targeting different IoT devices and expecting different responses, our honeypot should adaptively choose the responses which can better mislead the attackers to perform follow-up attacks. In HoneyIoT, 
we formulate the interactions between the honeypot and the attacker as a MDP and leverage reinforcement learning algorithms to select the proper responses.

In a typical MDP, the agent interacts with the outside world called environment through a series of actions. 
At each step, the agent observes the environment state and takes an action. As a result of the action, the environment makes corresponding changes and transits to another state. 
Meanwhile, the agent may receive a reward generated by the environment. 
The agent will continue this process, and get accumulated rewards after every action until the session is over. The agent seeks to maximize the accumulated rewards by taking proper actions. 

\begin{figure}[tbp]
    \vspace*{-0.3in} 
    \centering
    \includegraphics[width=7 cm]{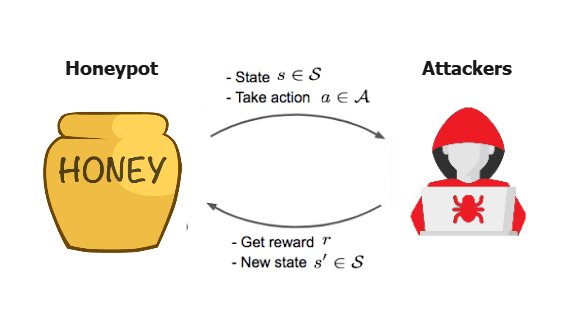}
    \vspace*{-0.4in} 
    \caption{Reinforcement learning model}
    \label{fig:RLModel}
    \vspace*{-0.3in} 
\end{figure}

As shown in Figure \ref{fig:RLModel}, in our case, the honeypot is the agent that interacts with the environment containing different attackers.
At each step, the honeypot agent observes the environment state by analyzing the attacker's request packet. 
The agent then takes an action to choose a proper response to fulfill the attacker's request. 
After receiving the response, the attacker may send another request that will cause the agent to enter a new state 
or terminate the session by stopping sending any requests. 
A reward is collected whenever the attacker tries to exploit a vulnerability, 
upload a malware or terminate the attack session. 
In the following subsection, we describe the state space, the action space, the state transition probabilities, 
and the reward function of our MDP model in detail.

\subsection{MDP Model Formulation} \label{sec:MDPModelDetail} 

\textbf{State Space.}
The state of MDP model should represent the current situation of the reinforcement learning agent. In our case, it should contain the interactions between the attacker and the honeypot.  
In order to clearly represent the attacker's session history, we define the state as a series of packets received from the attackers.  
Let $s_c$ denote the state space, and $s_{c}=H\left(p_{1}, p_{2}, \ldots, p_{c}\right)$,  
where
$s_c$ denotes the state space, and $p_i$ denotes the $i^{th}$ request packets received from the attacker.
However, directly using the packets from the attack trace may lead to sparse state space 
since the same type of request packets from different attackers may only vary slightly. 
To address this problem, some states may be aggregated, i.e., request packets with the same path and query strings are represented by a single state. 
In addition, we manually add a terminating state if no new packet is received from the attacker after a certain amount of time to indicate the end of the session.

\textbf{Action Space.}
Each response from the IoT device is labeled as a discrete action for the honeypot agent. 
That is, $a \in A = \{q_1, q_2, \cdots \}$, where $A$ is the action space which is a set of all possible responses from IoT devices, and $q_i$ denotes a specific response from certain IoT devices.
Since some IoT devices may not be able to respond to certain requests as they do not provide certain services, only a subset of actions is available at a given state in our MDP model. Therefore, the action space for our MDP model is essentially a multi-discrete action space. 
When a specific action is taken by the honeypot agent, the corresponding response will be rewritten (i.e. the date value) and forwarded to the attacker in order to provide a real-time high-fidelity response. The details on packet modification will be discussed in Section \ref{dfa}.

\textbf{State Transition Probabilities.}
In our model, the state transition probability (Equation \ref{equ:StateTransition}) can be described as a transition function $T \left(s, a, s^{\prime}\right)$ where $s$ is the current state of the environment, $a$ is the action taken by the agent, 
and $s'$ is the next state of the environment. 
In our case, the next state $s'= H\left(p_{1}, \ldots, p_{c}, p_{c+1}\right)$ refers to the packet received after taking action $a$ at the current state $s=H\left(p_{1}, \ldots, p_{c}\right)$.  
To better model the attacker's behavior, we utilize the attack trace collected through our system. We calculate the percentage of the occurrences of $(s,a,s')$ over the occurrences of all combinations containing the current state $s$ and action $a$. 
The percentage is then used as the state transition probability.
\begin{equation} \label{equ:StateTransition}
\begin{split}
T \left(s, a, s^{\prime}\right) & = P\left(S_{t}=s^{\prime} \mid S_{t-1}=s, a_{t}=a\right) \\
 & = \frac{C\left(s, a, s^{\prime}\right)}{ \sum\limits_{x \in S} C(s, a, x)}
\end{split}
\end{equation}
\begin{figure}[tbp]
    \centering
    \hspace{1cm}
    \includegraphics[width= 6cm]{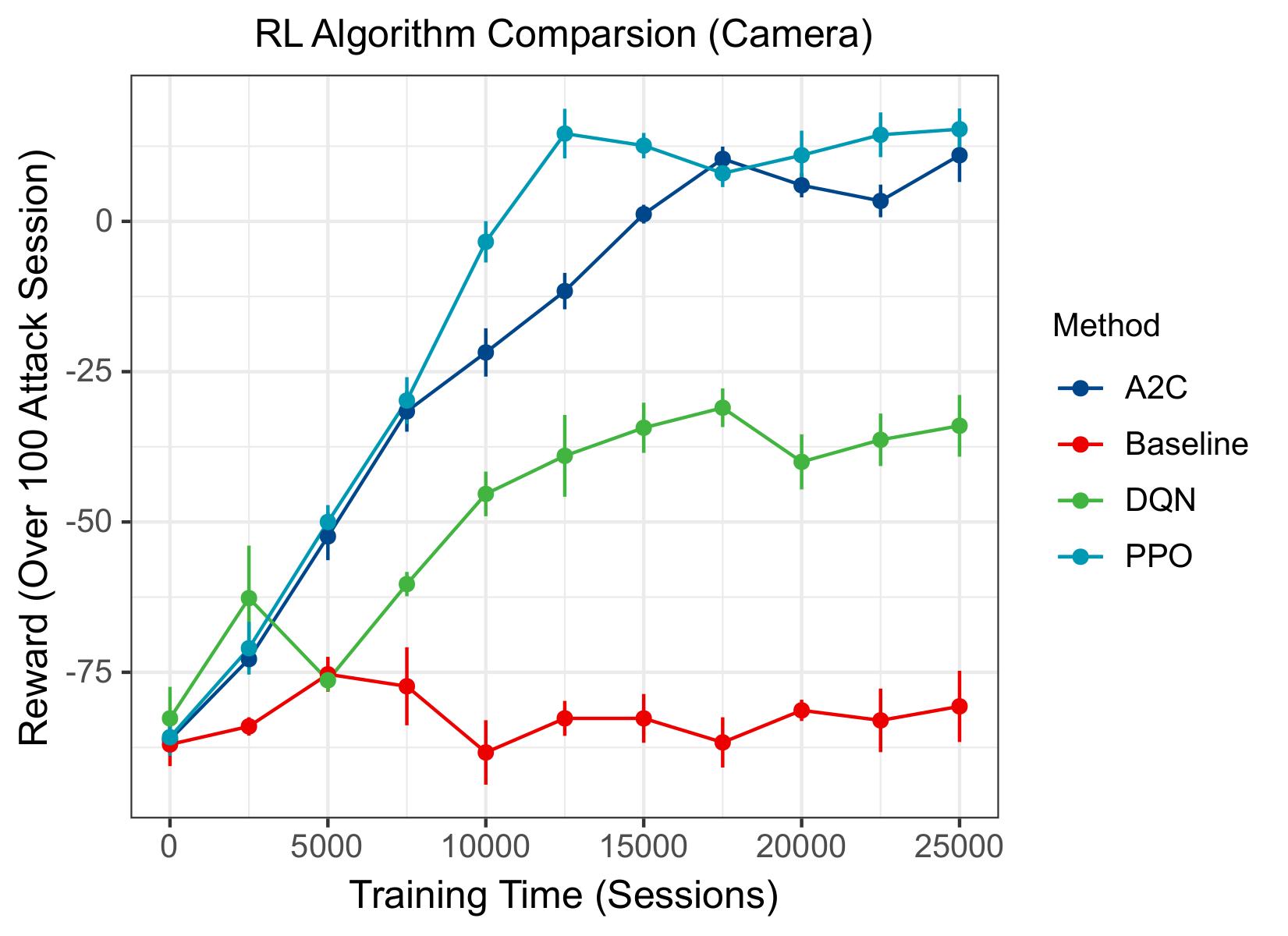}
    \vspace*{-0.15in}
    \caption{Comparison of different learning algorithms}
    \label{fig:cameraRL}
    \vspace*{-0.25in}
\end{figure}

\textbf{Reward Function.}
Since the ultimate goal of our honeypot is to mislead the attacker to upload the malicious code, we should in general reward the responses that lead to a real attack while punishing those that cause the attacker to end the current session. 
Typically, an attacker will exploit a vulnerability before uploading malware, so we assign a moderate reward for vulnerability exploitation, while reserving the highest reward for uploading malicious code. 
The intermediate reward reflects the potential progress we make when selecting a response in a given state. 

In our context, the reward function can be described as $R \left(s, a, s^{\prime}\right)$, where $s$ is the current state of the environment, $a$ is the action that our honeypot agent takes and $s'$ is the next state of the environment. 
In particular, we set the intermediate reward $R \left(s, a, s^{\prime}\right)$ for each valid state transition to 0 unless the last packet in the next state $s^{\prime}$ is vulnerability exploitation, malware upload or terminating state. If the attacker terminates the session without sending any malware download command or exploiting any vulnerabilities, we assign a negative reward. If any vulnerabilities are exploited, we add a positive intermediate reward. If the attacker uploads malware, we assign a large positive reward to this session. 

{\bf Algorithm Selection.} 
In recent years, many reinforcement learning algorithms such as Proximal Policy Optimization (PPO) \cite{PPO}, Deep Q Network (DQN) \cite{DQN} and Advantage Actor Critic (A2C) \cite{A2C} have been proposed.
In particular, PPO is a policy based method. A reinforcement learning policy is a mapping from the current environment observation to a probability distribution of the actions to be taken. In PPO, we build and update a policy to maximize the long-term reward while learning. 
DQN is a value based method where we store and update the value function for each state and action pair, and use the value function to deduce the best action at a given state.
A2C is a hybrid of policy based method and value based method. It consists of a policy based actor which controls 
how the agent acts, and a value based critic that measures the effectiveness of the agent's action.

In order to choose the most suitable algorithm for HoneyIoT,
we first turn the collected attack trace into the MDP model defined in section \ref{sec:MDPModelDetail} and then use the Open AI Gym environment \cite{brockman2016openai} 
and stable baseline3 \cite{StableBaseline} to test the effectiveness of different reinforcement learning algorithms. 
For comparison purposes, we also set up the baseline approach where no reinforcement learning algorithm is used. The baseline agent chooses each valid candidate response with equal probability.
In our setting, a positive reward of 5 and 1 is given when the agent receives a malware upload or vulnerability exploitation in an attack session, and a negative reward of -1 is given otherwise. 
In order to minimize the variance, we evaluate the reward over 100 random attack sessions multiple times for different RL algorithms with different training times. The results in Figure \ref{fig:cameraRL} show that PPO has the fastest convergence rate and the best average reward compared to other algorithms. Therefore, HoneyIoT uses PPO as the learning algorithm.
\vspace*{-0.1in}
\subsection{Differential Analysis based Content Mutation against Fingerprinting Attacks}
\label{dfa}

Although the reinforcement learning model can select a valid response when receiving the request, the attackers may still be able to detect the honeypot through fingerprinting attacks. 
Fingerprinting attacks \cite{Survey_Vuner,HoneypotDetection1,HoneypotDetection2,HoneypotDetection3} have long been an effective tool to detect and label honeypots.  
They have been widely applied in cyber attacks mainly because they can greatly improve the success rate of intrusion.
In a typical fingerprinting attack, the attacker first collects information from remote hosts 
through active probing or passive eavesdropping. 
The attacker then uses some heuristic rules \cite{fingerprintIoT3USENIX,fingerprintIoT4jsac} to match the responses with existing fingerprint database and analyzes the inconsistency. 
Most of the existing open-source honeypots are not effective against fingerprint attacks \cite{HoneypotDetection1} due to their open nature. For example, Dionaeca 
can be identified through its fixed banner information, Cowrie 
can be detected through its error messages and Glastopf can be identified by analyzing its HTTP responses.
Without proper content mutation, the attacker may identify the fixed replying logic of the remote host 
through fingerprinting attacks and detect the honeypot. 

To address these fingerprinting attacks, we need to provide responses that are unique to different attackers and follow the internal logic of IoT devices. We need to modify the packet selected by the RL model before returning it to the attacker. 
However, it is a challenge to locate the field that needs to be updated and figure out how we should update these values. 
As IoT devices from different manufactures have different internal logic, generating responses by reverse engineering each IoT device can be time-consuming. 
To address this problem, we apply differential analysis based method to analyze the responses and extract their mutation fields and mutation logic. 

Differential analysis has been applied to IoT security and privacy analysis domain as it can help researchers glean valuable insight by analyzing closely related inputs.
For example, IotSpotter \cite{Differential2CCS} uses differential analysis to identify IoT-specific libraries among different mobile apps. 
Continella {\em et al.} \cite{Differential1NDSS} proposed a privacy leak detection method for mobile apps by analyzing the difference among the network traces which are generated based on users' requests containing various private information (e.g. location). However, none of them has been applied to IoT honeypot design.

We identify the {\em mutation fields} by leveraging differential analysis techniques as follow. 
Based on the collected trace, we find responses generated by the same request (i.e., with the same request path and query string) to the same IoT device at different time. 
By comparing these responses, we find that some fields do not change (e.g., static web content) and some fields change (e.g.,  date, time, sensor readings). The fields that change are the {\em mutation fields}.  

Since different attackers may send the same request to the same IoT device at different time, the response from the IoT 
device can be directly used for differential analysis. For some rare request that is only sent by few attackers, 
we replay the attack request offline towards the IoT device multiple times to collect enough responses for differential analysis. We then extract the responses and use the Needleman-Wunsch algorithm \cite{DA_algorithm} to identify the mutation fields. 
These mutation fields can be classified into three categories based on the cause of the mutation:

\begin{itemize}[topsep=3pt]
    \item \textbf{Timing}:  The mutation field is affected by the current time of the IoT device. These mutation fields are usually in HTTP headers, device diagnostic logs, or user interface. 
    
    \item \textbf{System}:  The mutation field is determined by the system and physical status of the IoT device such as the pan and tilt angle of an IoT camera, the temperature value of a thermostat, the switch status (on, off) of a 
    smart plug, etc. 
   
    \item \textbf{Random}:  The mutation field does not have a determined value. These mutation fields may be session identifiers, encryption value, or random values caused by the software non-determinism.
\end{itemize}

For different categories of mutation fields, we have different rules on updating their values to provide high-fidelity responses. 
First, since the collected responses may be weeks before the real HoneyIoT deployment, 
we need to update the time related mutation fields to avoid inconsistency. 
In general, we generate some simple heuristic rules for time related requests to rewrite these mutation fields.
For example, we replace the date field of the HTTP header with the current server time and change the time related themes in the user interface accordingly.

Second, for system category, since the physical or system status of the IoT device may change over time, we need to provide different system values to the attackers at different time. The system value should be bounded by the physical capability of the IoT device (e.g., pan and tilt angle of a camera) or the environment (e.g., temperature of a thermostat). 
Since the values generated by the IoT devices in the past are valid, we store these valid system values 
in a database. At run time, HoneyIoT will randomly select a system value from the database to rewrite the corresponding
mutation field.

Third, for mutation fields that do not have a determined value, the attacker will not be able to detect the inconsistency 
as long as we replace it with a different random value. For HoneyIoT, we record the features (e.g., length, maximum, minimum) 
of these random values, and then 
generate random values based on these features for the corresponding mutation field. 

\begin{figure}[tbp]
    \vspace*{-0.1in}
    \centering
    \includegraphics[width=8.5cm]{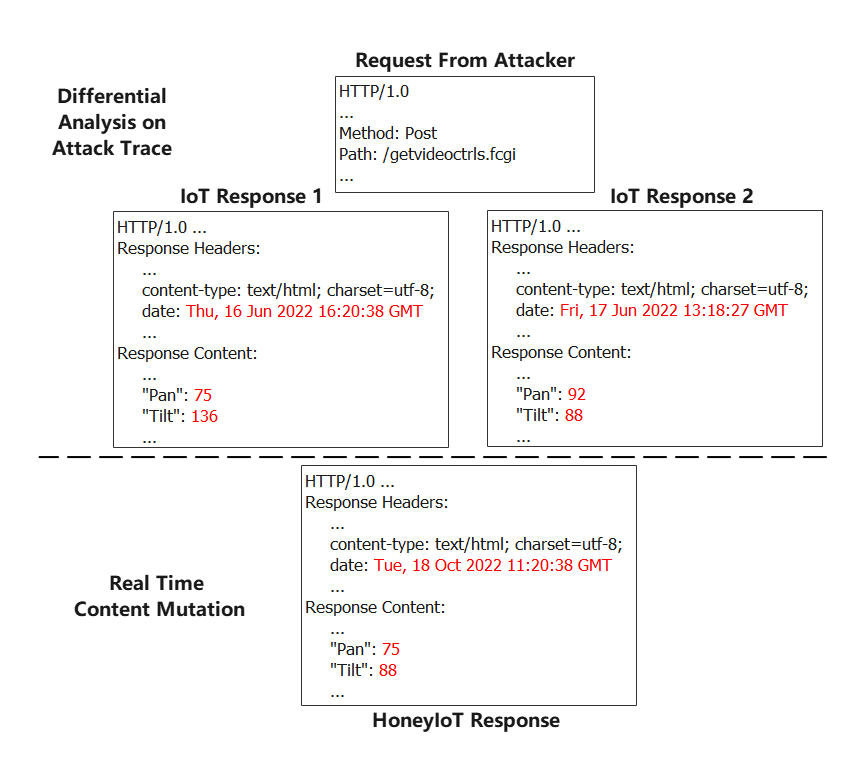}
    \vspace*{-0.4in}
    \caption{An example of using differential analysis to update the mutation field}
    \label{fig:DA_Example}
    \vspace*{-0.3in}
\end{figure}

Figure \ref{fig:DA_Example} shows how to update the mutation fields with a simple example 
based on attacks against an NC220 web camera. 
This "/getvideoctrls.fcgi" request allows the attacker to acquire video control related information, 
which is determined by the orientation of the camera. 
As shown in the figure, 
there are three mutation fields, time related mutation field ``date", system related fields ``Pan" and ``Tilt". 
For the time related field ``date", we rewrite it with the current time of the HoneyIoT server.
The ``Pan" and ``Tilt" fields represent the horizontal and vertical angles of the IoT camera. 
As they are related to the physical orientation of the device, we classify them as the system category and store them in a database. At run time, HoneyIoT selects 75 for ``Pan" and 88 for ``Tilt". 

By employing differential analysis based content mutation, 
HoneyIoT can effectively mitigate fingerprinting attacks and mislead attackers to launch followup attacks and upload malware. As future work, we aim to enhance the content mutation module by conducting more in-depth analysis of the correlations among different mutation fields, with the aim of generating more realistic and consistent responses that can better deceive evolving attackers.  

\vspace*{-0.1in}
\subsection{System Design}
\label{sec:HoneyIoTSystemDesign}

\begin{figure}[tbp]
    \centering
    \includegraphics[width=8cm]{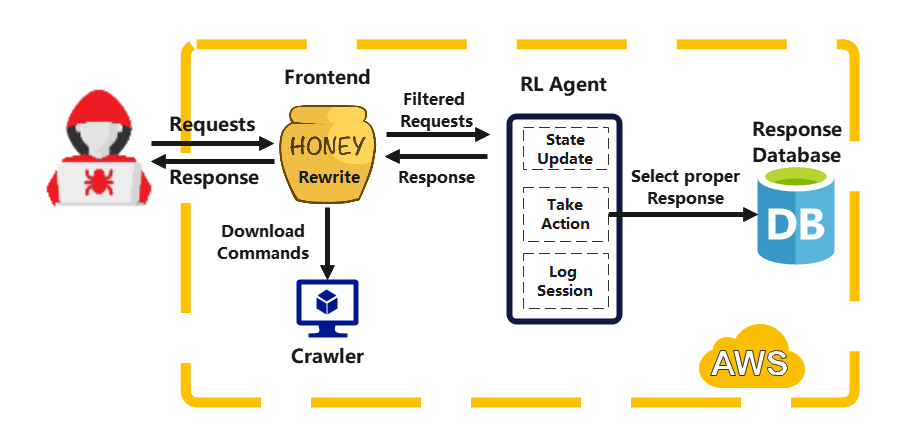}
        \vspace*{-0.25in}
    \caption{HoneyIoT system structure}
    \label{fig:HoneypotStructure}
        \vspace*{-0.2in}
\end{figure}

As shown in Figure \ref{fig:HoneypotStructure}, HoneyIoT consists of two main components: the frontend and the reinforcement learning agent.

The frontend is a virtual machine which opens some ports to provide services similar to a real IoT device. It needs to parse and analyze  attackers' request packets efficiently.
Once the frontend receives an attacker's request packet, it extracts the crucial attributes of the data (i.e. source IP, target port, request path, query string, etc) and forwards them to the reinforcement learning agent. 
If any malware downloading command (e.g., wget) is detected, the frontend forwards the command to a separate crawler in a sandbox. The crawler will parse the obtained command and establish a separate outbound connection to the attackers' Command and Control (C\&C) server and automatically collect malware for further analysis.
The frontend is also responsible for assembling and rewriting the response selected by the reinforcement learning agent. As discussed in the previous section, HoneyIoT will use different logic to mutate the value based on the type of mutation field in the response. For system value, HoneyIoT randomly selects a valid value from the database to rewrite the field. 
For timing value, HoneyIoT combines the session information and heuristic rules to generate the mutation value. 
For mutation field that does not have a determined value, 
HoneyIoT uses some random number and string generator to rewrite the value. 
Some HTTP header information will also be updated in order to ensure the fidelity of the packet. 

The reinforcement learning agent is responsible for emulating IoT devices by selecting a proper response based on the attacker's request. Upon receiving the attacker's request from the frontend, the reinforcement learning agent updates the state by appending the packet to the attacker's session history. It then takes an action by fetching a specific IoT device’s response from the response database following the reinforcement learning algorithm and updates the reward based on the reward function. During the interaction process, HoneyIoT also logs all attack traffic for further analysis. 

\begin{table*}[tbp]
\caption{Basic Statistics}
\vspace*{-0.1in}
\label{tab:basic}
\centering
{%
\begin{tabular}{|l|m{3cm}<{\centering}|m{2cm}<{\centering}|m{3cm}<{\centering}|}
\hline
                       & HoneyIoT-camera & Our Baseline & \begin{tabular}[c]{@{}c@{}}Existing Honeypot \\ (Snare \& Tanner)\end{tabular}\\ \hline
Attack Sessions        & 29467        & 26301 & 28660     \\ \hline
Average Session Length  & 7.57        & 4.97 & 4.02        \\ \hline
Vulnerability Exploit & 2963         & 843    & 731   \\ \hline
Malware Collected (Total)   &  467          & 92  &  14      \\ \hline
\end{tabular}%
\vspace{-0.1in}
}
\end{table*}

\section{Performance Evaluations}
\label{sec:eval}

In this section, we evaluate the performance of HoneyIoT and compare it to other IoT honeypots in various aspects.

\subsection{Experiment Setup}

We have implemented HoneyIoT and deployed it on the public Internet. Specifically, it runs on AWS using t2.small instance with 2G memory and one vCPU core. 
The frontend is mainly written in python and shell scripts. 
It opens ports identical to the real IoT devices it tries to emulate. For example, to emulate an IoT camera, the frontend 
opens the HTTP port 80, HTTPS port 443, RTSP port 554 and RTMP port 1935. We notice that different brands and models of IoT devices usually open some device specific ports providing services like mobile app control or firmware update. 
The frontend also opens these ports as there are attack traces collected over them. 

Upon receiving a request from an attacker, HoneyIoT processes the request and forwards it to the RL Agent. 
After the RL Agent chooses a valid response, the frontend rewrites the corresponding mutation field based on their categories (in Section \ref{dfa}).
We implement a separate crawler module to automatically collect malware from the attacker's control and command server once the frontend receives a downloading command. The malware sample is uploaded to VirusTotal\cite{VirusTotal} for malware classification and analysis.

In our implementation of the RL agent, we use stable baseline3 \cite{StableBaseline} to generate the RL model. 
As discussed in Section \ref{sec:MDPModelDetail}, we choose PPO as the RL algorithm in our model. We also implement a log module to store the interactions between the RL agent and the attackers for further analysis. The interaction log contains information such as the attacker's IP address, the request packets and reinforcement learning model's action at each step, etc.

We compare HoneyIoT with the following IoT honeypots: 

\begin{description}
\item{\bf Our Baseline:} It is identical to HoneyIoT except that it does not use any RL algorithm. Instead, we implement a simple heuristic algorithm to let our baseline honeypot select responses based on the MDP model directly. In the algorithm, each response has an equal probability to be chosen regardless of the attacker's session history.

\item {\bf Existing Honeypot:} We deploy an existing open-source honeypot called Snare \& Tanner \cite{Snare}, because it 
can effectively mimic the Web service of IoT devices opened on HTTP ports. Since the majority of the vulnerabilities of the emulated IoT devices are over HTTP ports, Snare \& Tanner can, at some degree, emulate these IoT devices. 
\end{description}

\subsection{Evaluation Results}

We trained the HoneyIoT agent using our collected attack trace and then deployed HoneyIoT, Our Baseline, and  Snare \& Tanner on AWS for two month (Oct. 2022- Dec. 2023) to evaluate their performance. 
In particular, we build HoneyIoT-camera based on the attack trace of five different cameras (as shown in Table ~\ref{tab:device}) and present the basic statistics in Section \ref{sec:comp_result}. 
Besides IoT camera, we also apply the idea of HoneyIoT to other IoT devices to test the extensibility of HoneyIoT. Specifically, we used the attack traces of six routers and five smart plugs to build HoneyIoT-router and HoneyIoT-smartplug and present their results in Section \ref{sec:Extensibility}.
In the remaining subsections, we evaluate the performance of HoneyIoT in terms of covertness and scalability. 

\subsubsection{Basic Statistics} \label{sec:comp_result}

As shown in Table \ref{tab:basic}, HoneyIoT-camera received 29,467 attack sessions, 2963 vulnerability exploitations, and 467 malware uploads from the attackers. Our baseline honeypot received 26,301 attack sessions, 843 vulnerability exploitations, 
and 92 malware uploads. Snare \& Tanner received 28,660 attack sessions, 731 vulnerability exploitations and collected 14 malware.
In order to estimate the capability of engaging attackers, we calculate the attack-session length by counting the request packets sent by attackers during each attack session. 
For better estimation, we excluded those sessions done by massive scanners which only perform a probe and then leave.  The average session length is 7.57, 4.97, and 4.02 for HoneyIoT-camera, our baseline honeypot and Snare \& Tanner, respectively.
These results indicate that HoneyIoT is more effective in engaging and misleading the attackers to upload their malicious code compared to our baseline and the existing honeypot.  

\begin{figure}[tbp]
    \centering
    \includegraphics[width= 8cm]{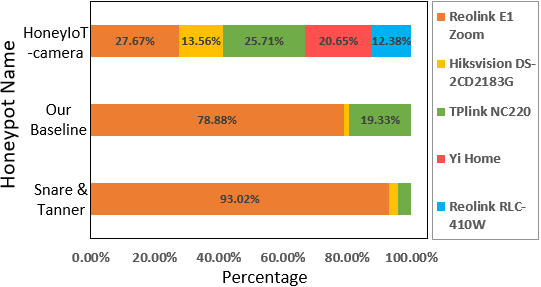}
        \vspace*{-0.1in}
    \caption{Vulnerability exploitation distribution}
    \label{fig:VulnerabilityComp}
        \vspace*{-0.2  in}
\end{figure}

Figure~\ref{fig:VulnerabilityComp} shows the vulnerability exploitation distribution of different honeypots. 
HoneyIoT-camera can successfully interact with the attackers and mislead them to exploit different vulnerabilities of all five IoT cameras, while our baseline honeypot and snare \& tanner only mislead attackers to the vulnerabilities of Reolink E1 zoom, TPlink NC220 and Hikvision cameras. 
In fact, most of the attacks collected by our baseline honeypot and snare \& tanner are through a vulnerability with CVE ID 2021-40149 on the Reolink E1 Zoom camera. 
It is a simple exploitation where attackers can obtain the SSL private key of the camera by launching a directory traversal attack. 
As shown in Figure \ref{fig:AttackGraphReolink}, compared to other sophisticated remote code execution attacks, this exploitation usually has a very short attack path and thus does not require pre-attack checks.
These results indicate that HoneyIoT-camera can effectively emulate the vulnerabilities of different IoT cameras and mislead attackers.

\begin{figure}[tbp]
    \centering
    \includegraphics[width= 6cm]{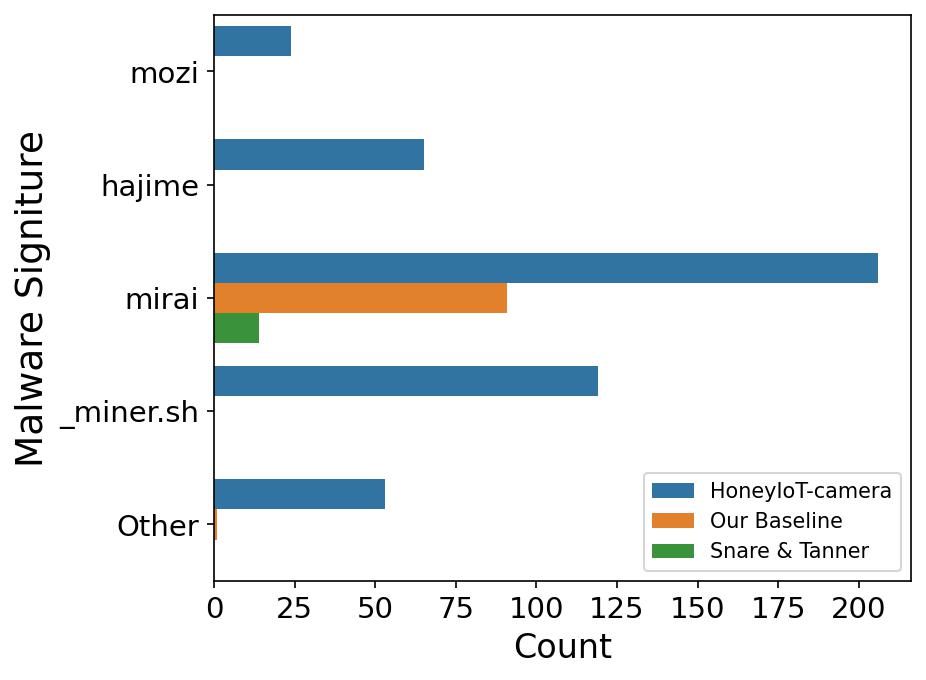}
        \vspace*{-0.15in}
    \caption{Malware uploaded}
    \label{fig:Mal_cam}
        \vspace*{-0.2in}
\end{figure}

Figure \ref{fig:Mal_cam} shows the malware collected by HoneyIoT-camera, our baseline honeypot, and snare \& tanner, which are identified through a malware analysis website called VirusTotal~\cite{VirusTotal}.  
Most of the malware collected by the HoneyIoT-camera can be classified as Mirai botnet and it's variants.  In particular, HoneyIoT-camera collected 467 botnet malware samples, where 24 are classified as Mozi \cite{Mozi}, 65 are classified as 
Hajime and 206 are classified as Mirai. Other than the botnet malware, HoneyIoT-camera also collected 119 miner malware that aim to mine cryptocurrency through the infected devices. 
Other 53 files collected from the attackers cannot be identified by VirusTotal, and thus we 
categorize them as ``other" malware. 
Although our baseline honeypot caught 92 malware and Snare \& Tanner caught 14 malware, they are mostly in the category of Mirai.
These results demonstrate that HoneyIoT can effectively mislead different attackers to upload their malicious codes.

\subsubsection{Extensibility} \label{sec:Extensibility}

In order to evaluate whether HoneyIoT can be extended for emulating other types of IoT devices besides IoT cameras, 
we trained reinforcement learning models (called HoneyIoT-router and HoneyIoT-smartplug) 
using the attack trace of six routers and six smart plugs (as shown in Table ~\ref{tab:device}) 
and deployed them on AWS. 
Within two months, the HoneyIoT-router received 40,531 attack sessions, 3675 vulnerability exploitation, collected 780  malware, and the average session length was 8.12.  HoneyIoT-smartplug received 22,104 attack sessions, 1973 vulnerability exploitation, collected 186 malware, and the average session length was 6.97.

\begin{figure} [tbp]
   \centering 
   \vspace{-0.1in}
   \subfigure[HoneyIoT-router]{ 
     \label{fig:Vul_router}  
     \includegraphics[height = 1.2 in]{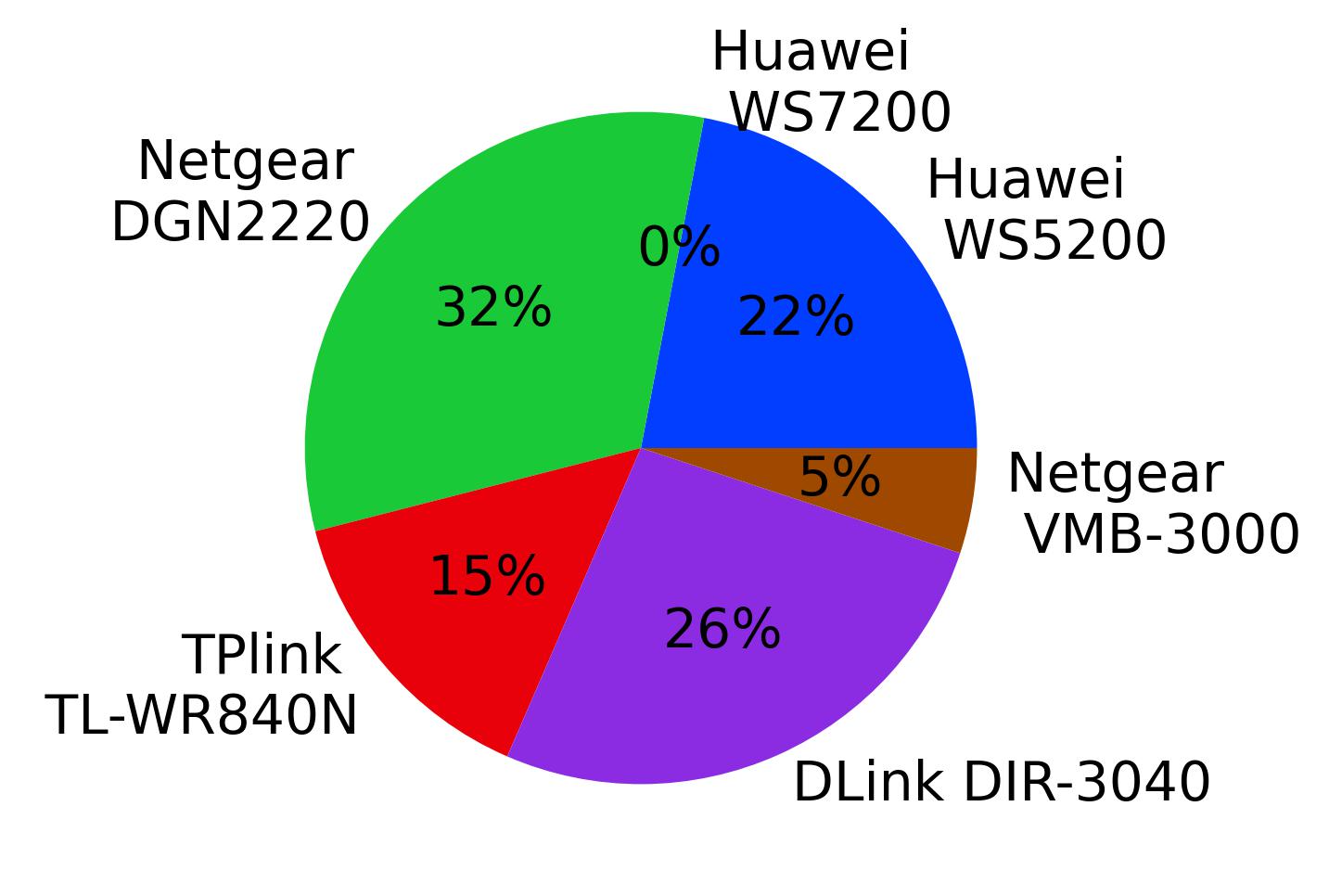}} 
   \subfigure[HoneyIoT-smartplug]{ 
     \label{fig:Vul_plug} 
     \includegraphics[height = 1.2 in]{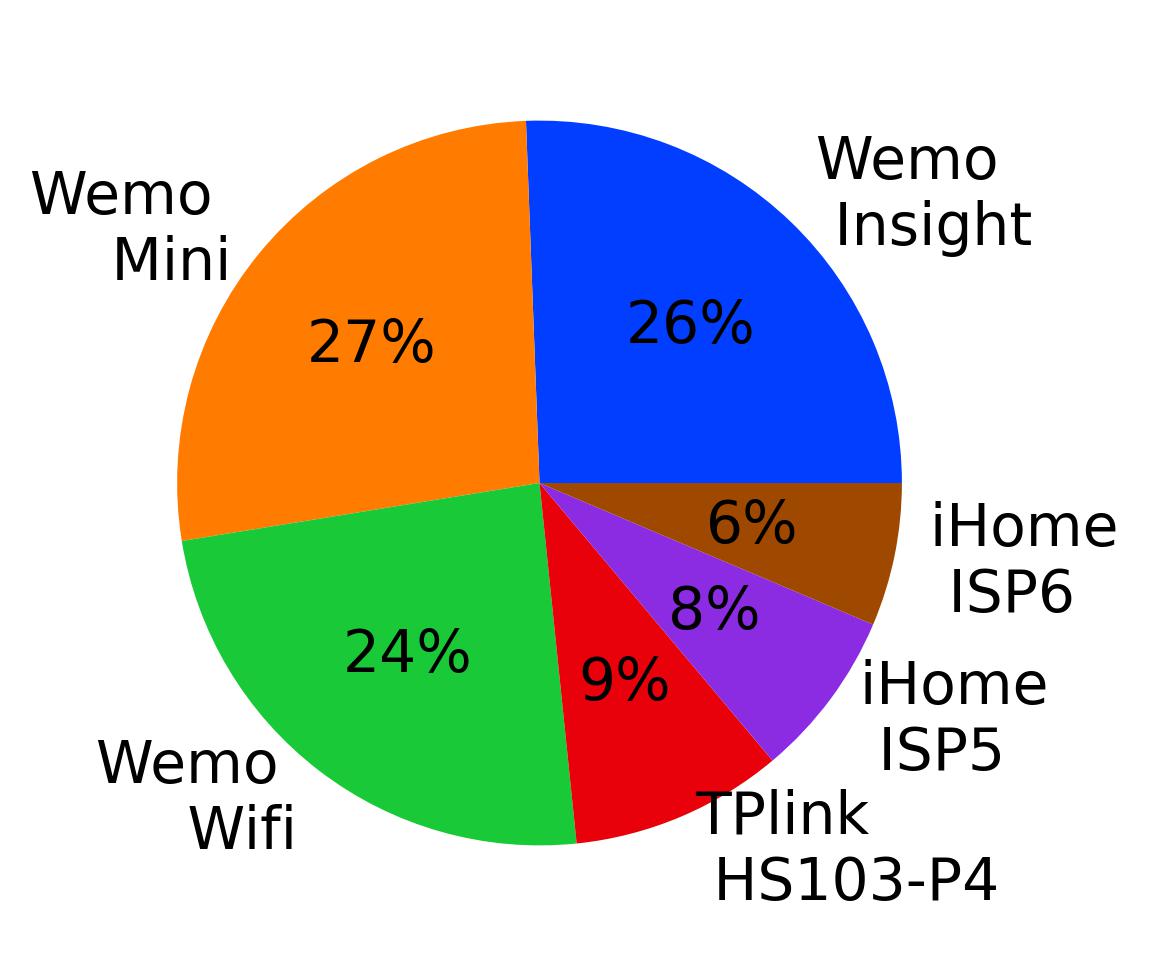}} 
    \vspace{-0.2in}
   \caption{Vulnerability exploitation distribution}
   \label{fig:VulnerabilityOther} 
   \vspace{-0.2in}
 \end{figure}

As shown in Figure \ref{fig:Vul_router}, HoneyIoT-router successfully misleads attackers to exploit the vulnerabilities of all emulated routers except Huawei WS7200, probably because it has no publicly exposed vulnerability.
After analyzing the log, we notice that the reinforcement learning model avoids using the response of Huawei WS7200 in most cases as it does not provide positive rewards.
On the other hand, HoneyIoT-smartplug successfully misleads attackers to exploit the vulnerabilities of all six simulated smart plugs as shown in Figure~\ref{fig:Vul_plug},

\begin{figure} [tbp]
   \centering 
   \subfigure[HoneyIoT-router]{ 
     \label{fig:Mal_router}  
     \includegraphics[width=1.62in]{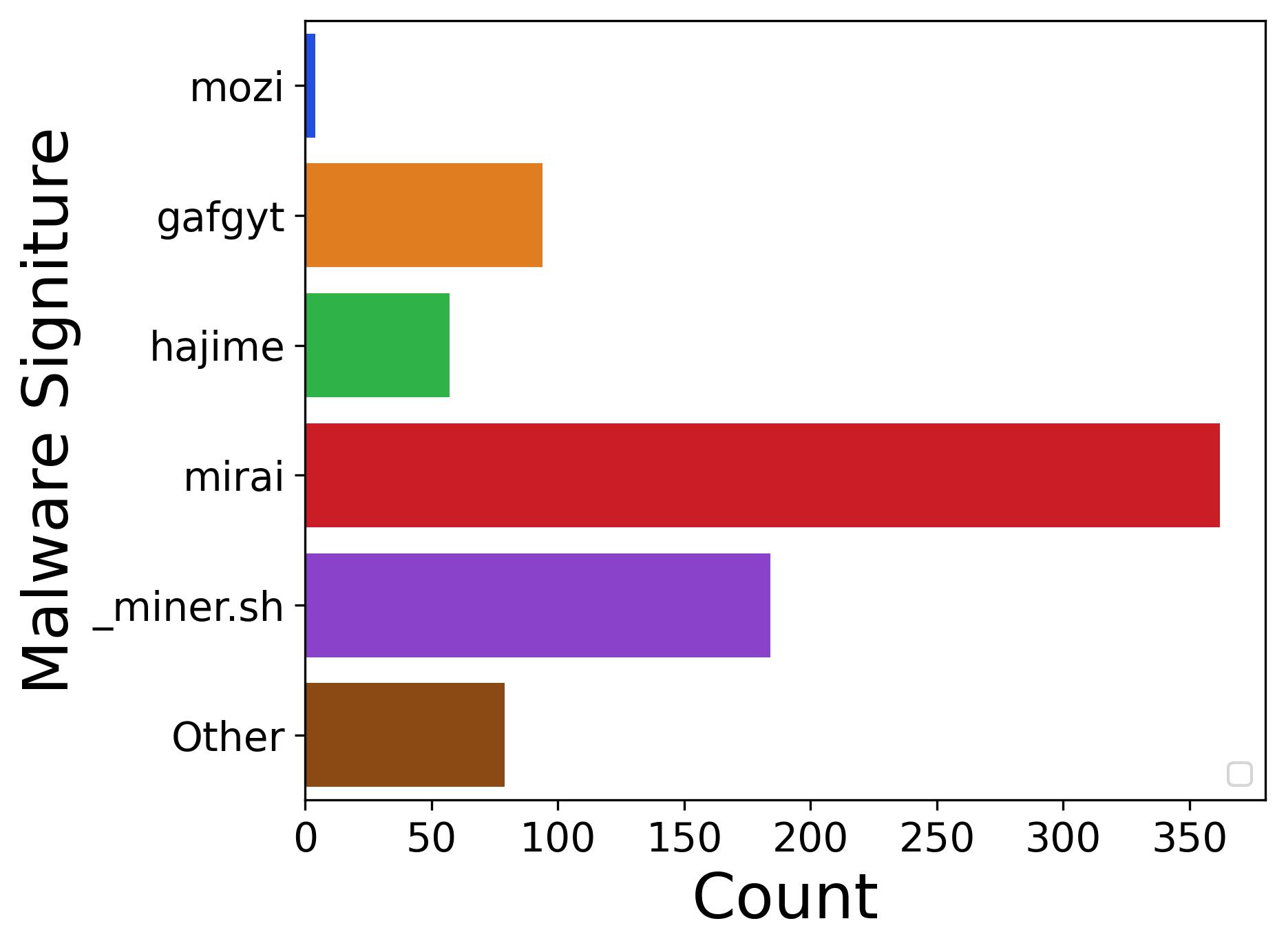}} 
   \subfigure[HoneyIoT-smartplug]{ 
     \label{fig:Mal_plug} 
     \includegraphics[width=1.62in]{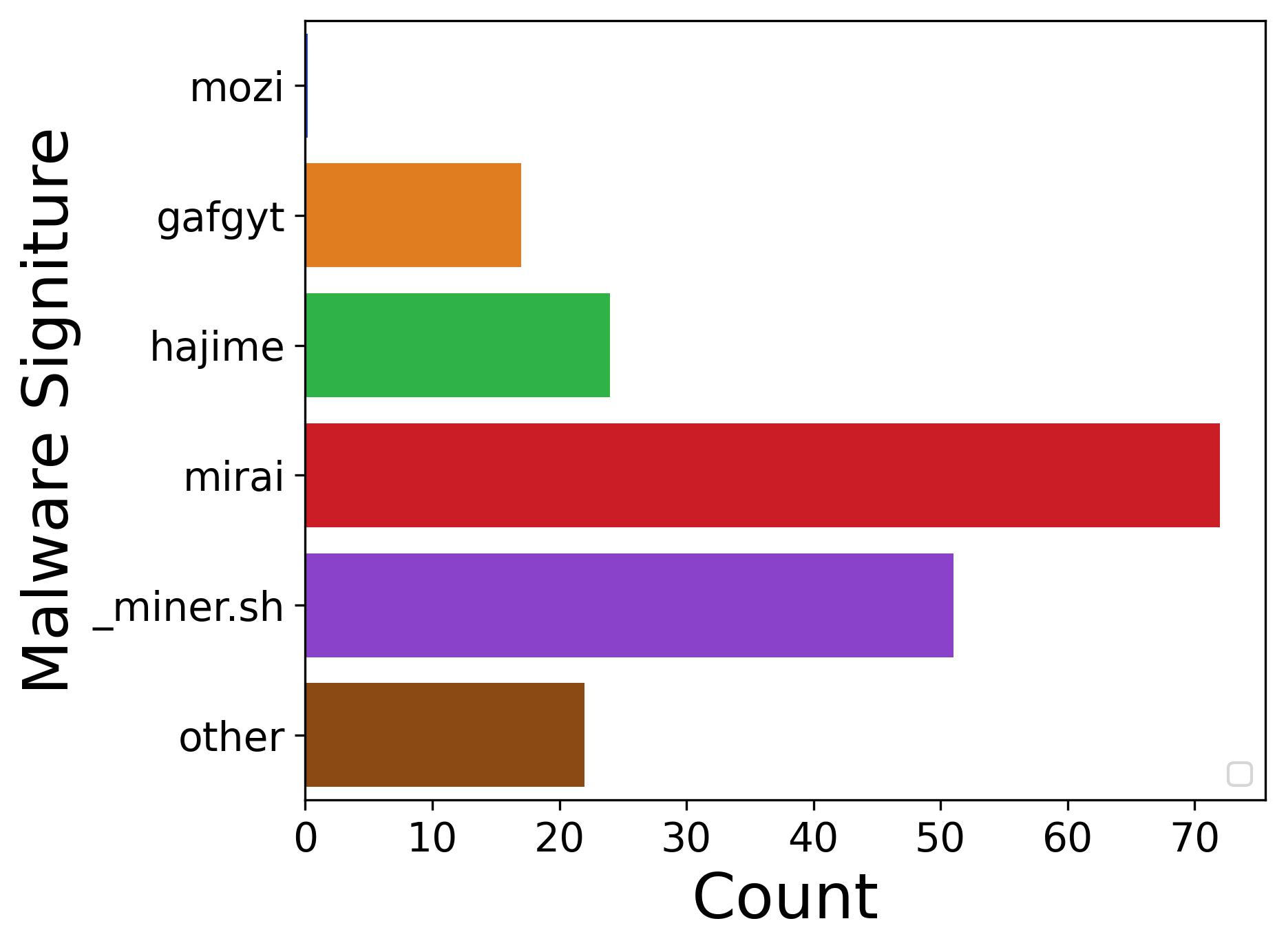}} 
    \vspace{-0.2in}
   \caption{Malware collected}
   \label{fig:MalwareOther} 
   \vspace{-0.15in}
 \end{figure}

Figure \ref{fig:Mal_router} shows the malware collected by HoneyIoT-router. For the 780 collected malware, 
4 are classified as mozi, 94 are classified as Gafgyt, 57 are classified as hajime, 362 are classified as Mirai, 184 are classified as miner, and 79 cannot be identified by VirusTotal. 
Figure \ref{fig:Mal_plug} shows the malware collected by HoneyIoT-smartplug, where 17 are classified as Gafgyt, 
24 are classified as hajime, 72 are classified as Mirai, 51 are classified as miner, and 22 cannot be identified by VirusTotal. 
Currently, HoneyIoT relies solely on VirusTotal to perform malware classification and analysis. Although most of the collected malware samples can be identified by VirusTotal, some files cannot be identified by it. 
Some of them are essentially empty files with randomly generated file names possibly due to the failure of attacker’s control and command server. Other files may represent new malware (e.g., 0-day malware) that require further investigation. 
These results together demonstrate that HoneyIoT can emulate the vulnerabilities of different routers and smart plugs, and can collect various types of malware from attackers. In other words, HoneyIoT can be extended to effectively emulate other types of IoT devices.

\begin{figure}[tbp]
    \centering
    \includegraphics[height = 3.5cm]{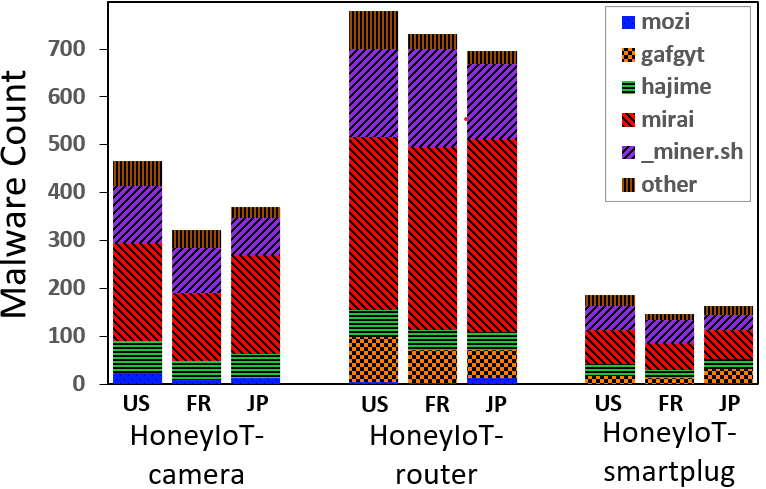}
    \vspace{-0.15in}
    \caption{Malware collected based on geographic location}
    \label{fig:GeoMalware}
    \vspace{-0.17in}
\end{figure}

We also deployed HoneyIoT-camera, HoneyIoT-router, and HoneyIoT -smartplug at three different locations (Virginia, Paris, Tokyo), and performed analysis on the collected malware based on the geographic locations.
As shown in Figure \ref{fig:GeoMalware}, HoneyIoT placed in US collected most malware regardless of the device it emulates. 
In particular, HoneyIoT-camera placed in US, France, and Japan collected 467, 322 and 370 malware,
HoneyIoT-router placed in the US, France, and Japan collected 780, 732 and 695 malware.
HoneyIoT-smartplug placed in the US, France, and Japan collected 186, 147 and 162 malware.
The performance difference among Honeypots deployed at different locations is mainly due to the behavioral difference among different groups of attackers. The attackers can hand-craft their mass-scanners to restrict the scanning area with the goal of reducing the overall scanning time and targeting specific devices that are widely used in a given region.

In addition, we notice that the types of malware collected by these three types of honeypots have certain degree of overlaps. This phenomenon indicates that some malware is independent of the compromised IoT device type. In other words, some attackers essentially treat IoT devices as general light-weight linux machines.
In this sense, for the purpose of malware collection, it is possible to obtain all types of popular malware 
over the public Internet just by emulating a potentially small subset of vulnerable IoT devices that are attractive to the attackers. In the future, we will further extend HoneyIoT to other types of IoT devices and try to locate this subset.

\subsubsection{Covertness} \label{sec:Covertness}

In this subsection, we evaluate whether HoneyIoT is covert against widely used reconnaissance and honeypot detection tools. 
The Shodan honeyscore\cite{shodan} is a well-known tool to check whether a remote host is a honeypot or not. 
Given an IP address, the Shodan honeyscore calculates the probability of the host to be a honeypot, in the range of 0.0 to 1.0, where 0.0 means that the host is definitively a real system, and 1.0 means that the host is a honeypot. 
The honeyscore is calculated by an undisclosed machine learning classification algorithm. 
According to Shodan, the honeyscore is affected by arguments such as the number of opened network ports, fingerprints of known honeypots, past records of the host IP address, and interactions with the host.
Therefore, we are interested in finding out the capability of HoneyIoT in dealing with this state-of-the-art reconnaissance tool.

\begin{figure}[tbp]
    \centering
    \includegraphics[ height = 3.5 cm]{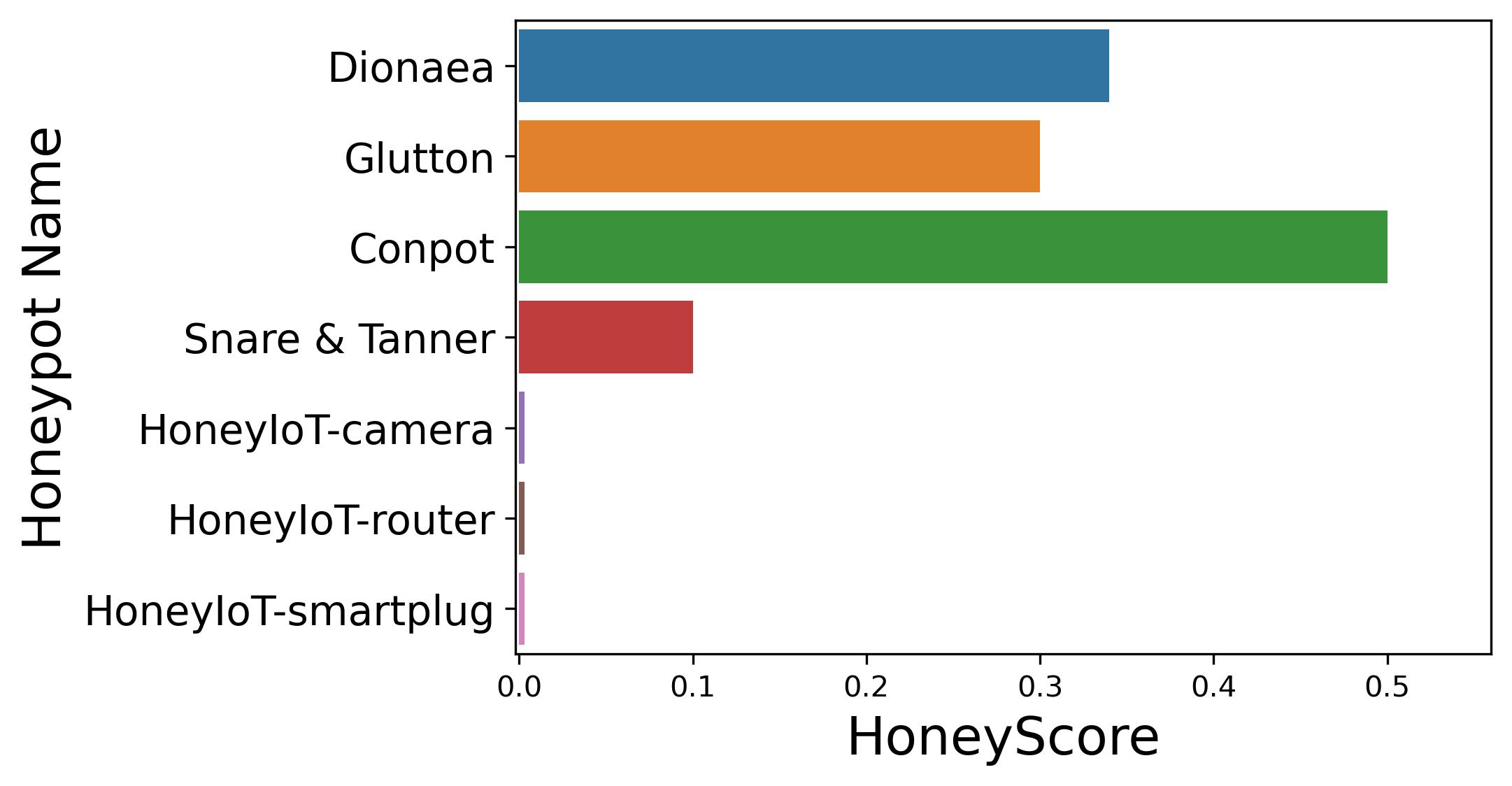}
    \vspace{-0.16in}
    \caption{Honey Score for existing honeypots and HoneyIoT }
    \label{fig:Shodan}
    \vspace{-0.18in}
\end{figure}

After the deployment of HoneyIoT on AWS, we wait for two weeks and then use Shodan API to acquire the honeyscores of HoneyIoT by providing their IP addresses. 
For comparison, we use T-pot \cite{T-Pot}, an all-in-one honeypot platform to deploy different open-source honeypots and collect their honeyscores. 
Specifically, Dionaea is a medium interaction honeypot that mimics multiple service including ftp, upnp and mqtt, Glutton emulates vulnerable ssh servers, and Conpot is designed for emulating Industrial Control System (ICS). 
In particular, we configure T-pot so that only one specific open-source honeypot runs at one time. 
To reduce variation errors, we deployed multiple instances for each type of honeypot on different locations for two weeks before collecting the honeyscores of these honeypots via Shodan API.

As shown in Figure \ref{fig:Shodan}, HoneyIoT has honeyscore of 0 which indicates that Shodan treats HoneyIoT as a real IoT device. On the other hand, the existing open-source honeypots all have honeyscores higher than 0. Dionaea, Glastopf, Conpot and Snare \& Tanner have honeyscores of 0.35, 0.3, 0.5 and 0.1.  
These results demonstrate that HoneyIoT is effective at maintaining covertness against state-of-the-art reconnaissance and honeypot detection tools.

\subsubsection{Scalability} \label{sec:Scalability}

Compared to real-device based honeypot which requires physical IoT devices \cite{guarnizo2017siphon,wang2020iotcmal} whenever the honeypot is active, HoneyIoT offers a more flexible approach.
Physical devices are only required during the attack trace collection phase, after which HoneyIoT can model IoT devices using the collected attack trace and use the trained model to interact with the attacker. 
Training the reinforcement learning model is a time-consuming but one-time job. In our case, it took us approximately two hours to process the attack trace and train HoneyIoT-camera with our server, which has an AMD Ryzen 7 5800 CPU and an RTX 3090 GPU.

At run time, the scalability and deployment cost of HoneyIoT depends mainly on its resource consumption. To evaluate the resource consumption of HoneyIoT, we collect the average CPU and memory usage of HoneyIoT in our AWS instance. 
In our experiment, we use the AWS T2.small instance which has one vCPU core and 2G memory. According to AWS, each vCPU core of a T2 instance is essentially a thread of a 3.3 GHz Intel Xeon Scalable processor.
For comparison, we also deploy several existing open-source honeypots including Dionaea, Conpot, and Snare \& Tanner on AWS using T2.small instance and monitor the CPU and memory usage of each honeypot process.

As shown in Table~\ref{tab:Scalability}, during the interactions with attackers, the average CPU usage of HoneyIoT-camera is 4.2\% and the average memory usage is 174 MB.
HoneyIoT-router and HoneyIoT-smartplug have similar average CPU and Memory usage. 
On the other hand, Snare \& Tanner has an average CPU usage of 7.1\% and an average memory usage of 282 MB.
Dionaea has an average CPU usage of 23.6\% and an average memory usage of 786 MB.
Conpot has an average CPU usage of 17.7\% and an average memory usage of 548 MB.
Dionaea and Conpot have much higher computational overhead than HoneyIoT mainly because they are not specifically 
designed for IoT devices and they emulate functions and services that are not supported by IoT devices. 
These results show that HoneyIoT is lightweight and has good scalability.

\begin{table}[]
\small
\vspace{0.05in}
\centering
\caption{Scalability}
\label{tab:Scalability}
\vspace{-0.1in}
\resizebox{.48\textwidth}{!}{%
\begin{tabular}{cccc}
\hline
\textbf{Honeypot} &
\textbf{\begin{tabular}[c]{@{}c@{}} Average \\ CPU usage \end{tabular}} &
\textbf{\begin{tabular}[c]{@{}c@{}} Average \\ Memory usage\end{tabular}} &
\textbf{AWS Instance}  \\ \hline
HoneyIoT-camera         &        4.2\%   &      174 MB              & T2.small  \\ \hline
HoneyIoT-router        &         4.7\%  &       187 MB         & T2.small  \\ \hline
HoneyIoT-smartplug          &    4.1\%    &     157 MB             & T2.small  \\ \hline
Dionaea   &    23.6\%    &      786 Mb             & T2.small     \\ \hline
Snare \& Tanner  &    7.1\%    &      282 Mb             & T2.small     \\ \hline
Conpot  &    17.7\%    &      548 Mb             & T2.small     \\ \hline

\end{tabular}%
}
\vspace{-0.2in}
\end{table}

\section{Related Work}
\label{sec:relatedwork}

There has been considerable research on using honeypots \cite{honeyproxy,SDNHoneyNet} to deceive the attackers. 
Many open-source or commercial honeypots have been deployed, especially for computer network services such as honeyd \cite{provos2003honeyd} and nepenthes \cite{baecher2006nepenthes}, etc. 
Recently, due to the wide adoption of IoT devices, IoT honeypots have also been 
developed to increase the overall security of IoT systems.

IoTPOT \cite{pa2015iotpot} is the first honeypot specifically designed for IoT devices. 
It is a low-interaction honeypot focusing on emulating telnet service which is used by many IoT devices.
Hakim {\em et al.} \cite{UPOT} introduced U-POT, an IoT honeypot framework specifically designed for the UPnP (Universal Plug and Play) protocol which is widely used in smart home such as surveillance cameras, smart bulbs, and smart switches. It uses device description files to automate honeypots and provide fake responses.
Seamus {\em et al.} \cite{RLHoney1} builds an SSH honeypot by implementing an interactive shell. They leverage reinforcement learning to conceal the honeypot characteristics.
In HoneyCam \cite{HoneyCam2021CNS}, to emulate the video streaming services of IoT cameras, the authors propose to prerecord 360\degree\ video and map the 360\degree\ video to different fields of view based on the attacker’s camera control commands.
These IoT honeypots mainly focused on emulating specific protocols or services that are widely used by the IoT devices. However, as discussed in section \ref{sec:background}, the attackers nowadays will perform various pre-attack checks 
to gather information for follow up attacks, and can notice that certain services are missing 
and suspect that they are not interacting with a real device.

As another approach, researchers leverage firmware images or real devices to build honeypots for IoT devices. 
Vetterl {\em et al.} proposed Honware \cite{Honware2019}, a virtual honeypot framework that can emulate different IoT devices based on their firmware image. 
HoneyCloud \cite{HoneyCloud} utilizes the firmware image of IoT devices to generate hardware and software IoT honeypots in order to collect attacks against Linux-based IoT devices.
Guarnizo {\em et al.} \cite{guarnizo2017siphon} proposed a high-interaction IoT honeypot architecture called SIPHON, which sets up honeypots on the cloud and leverages traffic forwarding technique to redirect attacker's commands back to IoT devices in their lab. 
Similarly, Tambe {\em et al.} \cite{VPNHoneypot} uses VPN to integrate the off-the-shelf IoT devices into a general honeypot architecture.
Luo {\em et al.} \cite{IoTCandyJar} proposed IoTCandyJar, an intelligent-interaction honeypot that emulates the request-response pattern of IoT devices. IoTCandyJar first acquires the attacker's request packets using low-interaction honeypots and probes the IoT devices on the public Internet to collect valid responses for the requests. 
IoTCMal \cite{wang2020iotcmal} uses real IoT devices to engage the attackers. It is a hybrid IoT honeypot which consists of a low interactive and a high interactive component. The low interactive component handles the login process and forwards the attacker's commands to the high interactive component that uses real IoT devices. Then, the attackers are essentially interacting with real IoT devices after the login phase.
However, most of the IoT device manufacturers do not publicize the firmware images of their products. In addition, always forwarding the attacker traffic to real IoT devices suffers from scalability issues and cannot be deployed at a large scale.   

Different from the aforementioned existing research, HoneyIoT does not limit itself to emulating specific protocol
or specific IoT device. Moreover, through reinforcement learning, HoneyIoT can adaptively interact with the attackers and mislead them to upload malware.

\vspace*{-0.3cm}
\section{Conclusions}
\label{sec:Conclusion}

In this paper, we proposed an adaptive high-interaction honeypot for IoT devices, called HoneyIoT,  which can better engage attackers leveraging reinforcement learning techniques.  
To learn how attackers interact with IoT devices, we first build a system for collecting attack traces from real devices. We model the attack behavior using a Markov decision process and use reinforcement learning to determine the best responses to engage attackers based on the collected traces. We also use differential analysis techniques to mutate response values in some fields to mitigate fingerprinting attacks. HoneyIoT has been deployed on the public Internet, and experimental results demonstrate its effectiveness in evading reconnaissance and honeypot detection tools, as well as its ability to mislead attackers into uploading malware.

In the future, we will enhance HoneyIoT's deception capabilities by developing more sophisticated reward functions and generating responses that are more consistent and realistic. 
Additionally, we will extend HoneyIoT to emulate a wide variety of IoT devices and deploy them across various public cloud platforms and enterprise networks to collect more malware samples and analyze attacker behaviors. Our ultimate goal is to leverage the collected malware samples to detect zero-day attacks before they can be widely launched. 

\section*{Acknowledgement}

This research was partially sponsored by the U.S. Army Combat Capabilities Development
Command Army Research Laboratory and was accomplished
under Cooperative Agreement Number W911NF-13-2-0045 (ARL
Cyber Security CRA).

{
\bibliographystyle{unsrt}
\balance
\bibliography{main}
\balance

}

\end{document}